\documentclass[12pt]{iopart}
\usepackage{graphicx}
\usepackage{epsfig} 

\begin{document}
\title[On O-X mode conversion in 2D inhomogeneous plasma...]
{On O-X mode conversion in 2D inhomogeneous plasma with a sheared
magnetic field}

\author{A Yu Popov}

\address{Ioffe Institute, St.Petersburg,
Russia} \ead{a.popov@mail.ioffe.ru}

\begin{abstract} The conversion of an ordinary wave to an extraordinary wave
in a 2D inhomogeneous slab model of the plasma confined by a sheared
magnetic field is studied analytically.
\end{abstract}

\section{Introduction}
The linear conversion of waves in nonuniform media is a common
phenomenon in physics and has been studied in fields as diverse as
RF heating of fusion plasmas, ionospheric physics and so on. The
particular case of the linear mode conversion, namely, the
conversion of an ordinary (O) wave to an extraordinary (X) wave at
electron cyclotron frequencies being a primary consideration for an
auxiliary electron heating and current drive in fusion devices was
intensively studied for plasmas with a cold plasma dielectric
tensor. A quantitatively accurate description of this problem was
paid much attention to a couple of decades ago within both the WKB
approximation and the full wave analysis for the 1D inhomogeneous
slab model of the plasma with zero shear~\cite{Ref1}-~\cite{Ref5}.
Having been renewed by Wendelstein 7-AS (W7-AS) considerable success
in realization of the mode conversion scheme~\cite{Laqua} the
problem was examined in a number of
papers~\cite{Weitzner},~\cite{Gospodchikov2D},~\cite{Irzak} within
the 2D plasma model and in~\cite{Gospodchikov3D} within the 3D
plasma model by analysis of a set of the reduced wave equations
valid in a vicinity of the O-X mode conversion layer. In all these
papers the curvature of the magnetic flux surfaces was neglected
under an assumption of high localization of the conversion region
and magnetic field was considered to be shearless with the straight
magnetic field lines. No effort was mounted to describe the O-X mode
conversion in more realistic case of the 2D inhomogeneous plasma in
the magnetic field with a shear except possibly the
paper~\cite{Lashmore-Davies} where the effect of the shear on the
efficiency of the O-X mode conversion was studied within the 1D
inhomogeneous slab model of the plasma. Here we are concerned with
the analysis of the O-X mode conversion problem for the 2D
inhomogeneous plasma with the sheared magnetic field.

\section{Introduction}
The linear conversion of waves in nonuniform media is a common
phenomenon in physics and has been studied in fields as diverse as
RF heating of fusion plasmas, ionospheric physics and so on. The
particular case of the linear mode conversion, namely, the
conversion of an ordinary (O) wave to an extraordinary (X) wave at
electron cyclotron frequencies being a primary consideration for an
auxiliary electron heating and current drive in fusion devices was
intensively studied for plasmas with a cold plasma dielectric
tensor. A quantitatively accurate description of this problem was
paid much attention to a couple of decades ago within both the WKB
approximation and the full wave analysis for a slab model of the 1D
inhomogeneous plasma with zero magnetic field
shear~\cite{Ref1}-~\cite{Ref5}. Having been renewed by Wendelstein
7-AS (W7-AS) considerable success in experimental realization of the
O-X mode conversion scheme~\cite{Laqua} the problem was examined in
a number of papers~\cite{Weitzner}-~\cite{Gospodchikov2Dc} within
the 2D plasma model and in~\cite{Gospodchikov3D} within the 3D
plasma model by analysis of a set of the reduced wave equations
valid in a vicinity of the O-X mode conversion layer. In all these
papers the curvature of the magnetic flux surfaces was neglected
under an assumption of high localization of the conversion region
and magnetic field was considered to be shearless with the straight
magnetic field lines. No effort was mounted to describe the O-X mode
conversion in more realistic case of the 2D inhomogeneous plasma in
a sheared magnetic field except possibly the
paper~\cite{Lashmore-Davies} where the effect of the shear on the
efficiency of the O-X mode conversion was studied within the 1D
inhomogeneous slab model of plasma. In this work we are concerned
with the analysis of the O-X mode conversion problem for the 2D
inhomogeneous plasma with the sheared magnetic field.
\section{Basic equations}
In this paper we consider plasmas with 2D inhomogeneity that assumes
the surfaces $n=const$ and $B=const$ being not approximately
parallel and the external sheared magnetic field. We treat a problem
in the simplest slab geometry accounting for plasma axisymmetry,
expecting that the results being obtained within this simple model
hold in toroidal plasmas.

Let us start with introduction of the Cartesian co-ordinate system
($x, y, z$) with its origin located at the O-mode cut-off surface
and coinciding with the center of an incident beam of the
monochromatic ordinary waves. The co-ordinates $x$, $y$ and $z$
being scaled in $c/\omega$, where $\omega$ is a frequency of the
wave and $c$ stands for the speed of light, imitate the flux
function, the poloidal angle and the toroidal angle, respectively.
We define the magnetic field as
\begin{eqnarray}
{\bf B}=B_0\left(\sin{\theta}{\bf e}_y+\cos{\theta}{\bf
e}_z\right),\nonumber
\end{eqnarray}
where $\theta=\theta_0+\delta\theta$, $\theta_0$ is an angle between
${\bf e}_\parallel={\bf B}/B_0$ and ${\bf e}_z$ at $x=0$ and
$\delta\theta=\delta\theta(x)$ ($\delta\theta=0$ at $x=0$)
originated due to the shear of the magnetic field. In this study we
assume for simplicity that $\theta_0=0$ and ${\bf e}_\parallel
\|{\bf e}_z$ at $x=0$. The parameters of inhomogeneity are assumed
to be small $|\nabla \ln{(n)}|\ll 1$ and $|\nabla \ln{(B)}|\ll 1$ as
in the previous studies~\cite{Weitzner}-~\cite{Gospodchikov3D}. In
such a case the WKB approximation is applicable for description of
the interacting waves except a narrow localized conversion layer in
a vicinity of the O-mode cut-off surface where the full-wave
analysis connecting the incoming and outgoing WKB waves is needed.
The WKB approach can give us not only the boundary conditions but
some hints on how to reduce the full-wave equations in the
conversion layer as well. The dispersion curves of the O and X modes
are adequately described by the cold plasma model. These curves are
coupled if the local wave-vector ${\bf n}$ with components $n_x={\bf
n}\cdot{\bf e}_x$, $n_\eta={\bf n}\cdot\left({\bf e}_\parallel\times
{\bf e}_x\right)$, $n_\parallel ={\bf n}\cdot{\bf e}_\parallel$  has
at the O-mode cut-off surface the only parallel component equal to
$n_{res}=\left(1+\omega/\omega_{ce}\right)^{-1/2}$ where
$\omega_{ce}$ is the electron cyclotron frequency at $x=0$, $y=0$.
This means the conversion of the O wave into the X wave  occurs if
in a close vicinity of the O-mode cutoff surface $n_\parallel\approx
n_{res}+\delta n_\parallel$ where $\delta n_\parallel\ll 1$ and $n_x
\ll 1$, $n_\eta\ll 1$. We expand the components of the cold
dielectric tensor and the angle $\theta$ near the origin of the
co-ordinate system into the Taylor serious up to the first-order
terms:
\begin{eqnarray}
\varepsilon_+-n_{res}^2\simeq-|\nabla\varepsilon_+|\left(x\sin{\varphi}-y\cos{\varphi}\right),~
\varepsilon_\parallel\simeq -|\nabla\varepsilon_\parallel|x,~
\theta\simeq Gx,\nonumber
\end{eqnarray}
where $\varphi$ is an angle between two surfaces
$\varepsilon_+-n_{res}^2=0$ and $\varepsilon_\parallel =0$, $G^{-1}$
is a length describing the magnetic field shear. The wave vector in
the conversion layer can be interpreted as the differential operator
\begin{eqnarray}
n_x=-i\partial_x,~n_\eta=-i\partial_y+Gx,~n_\parallel-n_{res}=i\partial_z,\partial_r=\partial/\partial
r\nonumber
\end{eqnarray}
acting on an appropriate component of the electric field. Using the
standard expansion procedure over the small parameters and keeping
the first order terms one can obtain the following set of equations
for the electric field components
\begin{eqnarray}
n_{res}D_+E_z-i\left(|\nabla\varepsilon_+|\left(x\cos{\varphi}-y\sin{\varphi}\right)-i2n_{res}\partial_z\right)E_+=0\nonumber\\
i2|\nabla\varepsilon_\parallel|xE_z-n_{res}D_-E_+=0\\
E_-=0,\nonumber
\end{eqnarray}
where $E_\pm=E_x\pm iE_y$, $D_\pm=\partial_\pm\mp n_{res}Gx$,
$\partial_\pm=\partial_x\pm i\partial_y$. Variation of the electric
field components in (1) is neglected as it results in the effects
being higher order of vanishing. For any detail of (1) derivation we
can refer readers to~\cite{Irzak b}. This set of equations is
different from that studied in~\cite{Irzak b}, which describes the
2D O-X mode conversion but misses the term $n_{res}Gx$ originated
due to the magnetic field shear.

This is our basic system which as it stands lost symmetry inherent
in the case of the magnetic field with zero shear $G=0$. It is at
first disturbing us by virtue of the fact that no separation of
variables and no existence of eigenvalues are permitted in the case
of the sheared magnetic field. Fortunately, this initial impression
is misleading as will be shown in the forthcoming sections. With the
use of the procedure proposed in~\cite{Irzak a} we will obtain the
second order partial differential equation instead of the system
(1). Then, applying an integral representation of Laplace integral
type as in~\cite{Irzak b} of unknown function, we will separate
variables and obtain eigenvalues and eigenfunctions of this partial
differential equation.

\section{Analysis of the system (1)}
Following a procedure having been proposed in~\cite{Irzak a} opens a
way of overcoming underlying mathematical difficulties and permits
us to separate variables in (1). As we assume an axisymmetric
plasma, we can seek a solution of (1) in a form $\sim\exp{\left(i
n_zz\right)},$ where $n_z=const$. Introducing new notations
\begin{eqnarray}
a=\left(2|\nabla\varepsilon_\parallel|/|\nabla\varepsilon_+|\right)^{1/4}E_z,\\
b=\left(|\nabla\varepsilon_+|/\left(2|\nabla\varepsilon_\parallel|\right)\right)^{1/4}E_+,\nonumber\\
g=G
n_{res}^2/\left(2|\nabla\varepsilon_\parallel||\nabla\varepsilon_+|\right)^{1/2},\nonumber\\
y-2n_{res}n_z/\sin{\varphi}/|\nabla\varepsilon_+|\rightarrow
y,\nonumber\\
x,y\cdot
\left(2|\nabla\varepsilon_\parallel||\nabla\varepsilon_+|\right)^{1/4}/n_{res}^{1/2}\rightarrow
x,y\nonumber
\end{eqnarray}
we read (1) as
\begin{eqnarray}
iD_+a+\left(x\cos{\varphi}-y\sin{\varphi}\right)b=0\\
iD_-b+x\cdot a=0\nonumber
\end{eqnarray}
where $D_\pm=\partial_\pm\mp g x$ and $\partial_\pm=\partial_x\pm
i\partial_ y$. Let us use a generalized Laplace transforms of
$a(x,y)\rightarrow a(x,q)=\hat{L}\left[a(x,y)\right]$ and
$b(x,y)\rightarrow b(x,q)=\hat{L}\left[b(x,y)\right]$ where
\begin{eqnarray}
\hat{L}\left[\,\cdot\,\right]=\exp{\left(-i\frac{g}{\sin{\varphi}}
q^2\right)}\int_C{dy\exp{\left(-iqy-i\frac{
\sin{\varphi}}{4g}y^2\right)}\left[\,\cdot\,\right]}
\end{eqnarray}
As we will see later, a contour $C$ can be chosen along the real
axis of the complex variable $q$. We introduce
$k=-2g/\sin{\varphi}\cdot q$ and seek $a(x,k),$ $b(x,k)$ in the form
\begin{eqnarray}
a(x,k)=\hat{\partial}_-f-\left(\left(g-i\beta\cos{\varphi}\right)x+i\beta\sin{\varphi}k+2\beta
g\partial_k\right)\cdot f,\\
b(x,k)=\beta\cdot\hat{\partial}_+f+\left(g\beta + i \right)x\cdot
f,\nonumber
\end{eqnarray}
where $\hat{\partial}_\pm = \partial_x \pm i
\partial_k$, $\beta=\left(\xi
-ig\right)\exp{\left(i\varphi\right)}$,
$\xi=\left(\cos{\varphi}-g^2-i
\sin{\varphi}\right)^{1/2}=|\xi|\exp\left(-i\psi\right)$,
$|\xi|=\left(1-2g^2\cos{\varphi}+g^4\right)^{1/4}$,
$2\psi=\arctan{\left[\sin{\varphi}/\left(\cos{\varphi}-g^2\right)\right]}$.
Using (4) and (5) we derived a partial differential equation of
second order for a function $f$ instead of the system (3) of two
partial differential equations of first order for $a$ and $b$
\begin{eqnarray}
\partial_x^2f+\partial_k^2f+\left((\cos{\varphi}-g^2)x^2-\sin{\varphi}\cdot
xk+i|\xi|\exp{\left(-i\psi\right)}\right)f=0
\end{eqnarray}
This equation deserves few comments. For zero shear, $g=0$, (6)
coincides with the equation analyzed in~\cite{Irzak a}. For the
sheared magnetic field we estimate the parameter $g$ as $g\sim
L_n/L_0\ll 1$, where $L_n$ is a scale  on which the the density
profile at the O-mode cut-off surface changes and $L_0$ is a minor
radius. Henceforth, we keep in mind that the coefficient
$\cos{\varphi}-g^2$ at $x^2$ in (6) is positive. At $\varphi=0$ we
return to the equation treated within the 1D O-X mode conversion
problem in the sheared magnetic field. First who analyzed this case
by the WKB analysis were Cairns and
Lashmore-Davies~\cite{Lashmore-Davies}. To separate variables in (6)
we introduce
\begin{eqnarray}
u=\sqrt{|\xi|}\left(\cos{\psi}\cdot x-\sin{\psi}\cdot k\right)\nonumber\\
v=\sqrt{|\xi|}\left(\sin{\psi}\cdot x+\cos{\psi}\cdot
k\right)\nonumber
\end{eqnarray}
that yields
\begin{eqnarray}
\partial_u^2f+\partial_v^2f+\left(\cos{\psi}^2u^2-\sin{\psi}^2v^2+i\exp{\left(-i\psi\right)}\right)f=0
\end{eqnarray}
A particular solution of (7) is
\begin{eqnarray}
f(u,v)=\sum_{p=0}^\infty{c_pD_{i\nu_p/\pi}\left(u\right)\phi_p\left(v\right)},
\end{eqnarray}
where
$\nu_p=\pi|\tan{\psi}|\left(p+\left(1-sign{(\varphi)}\right)/2\right)$,
$D_{i\nu_p/\pi}(u)\equiv
D_{i\nu_p/\pi}\left(\sqrt{2\cos{\psi}}\exp{\left(i\pi/4\right)}u\right)$
is the parabolic cylinder function~\cite{Bateman},
$\phi_p(v)\equiv\phi_p\left(\sqrt{\sin{|\psi|}}v\right)$ are the
Hermitian polynomials
\begin{eqnarray}
\phi_p\left(v\right)=\left(2^p\sqrt{\pi}p!\right)^{-1/2}
\exp{\left(-v^2/2\right)}H_p\left(v\right)\nonumber
\end{eqnarray}
possessing a property
$\int{\phi_p\left(v\right)\phi_k\left(v\right)dv}=\delta_{pk}$.
Substituting (8) in (5) we obtain
\begin{eqnarray}
a(x,k)=R\left(\exp(i\psi)|\xi|+ig\right)-I\left(\exp(i\psi)|\xi|-ig\right)\\
b(x,k)=R+I,\nonumber
\end{eqnarray}
where
\begin{eqnarray}
I=\sqrt{|\xi|}\sum_{p=0}^\infty c_p D_{i\nu_p/\pi}\left(u\right)
\left(\partial_v+\sin{\psi}\cdot v\right)\phi_p\left(v\right),\\
R=\sqrt{|\xi|}\sum_{p=0}^\infty c_p\phi_p\left(v\right)
\left(-i\partial_u+\cos{\psi}\cdot
u\right)D_{i\nu_p/\pi}\left(u\right).
\end{eqnarray}
Simple but tedious algebra yields
\begin{eqnarray}
I=\sqrt{|\sin(\psi)|}\sum_{p=0}^\infty C_p
D_{i\gamma_p/\pi}\left(u\right)
\phi_p\left(v\right),\\
R=\sqrt{i\cos(\psi)}\sum_{p=0}^\infty C_p
D_{i\gamma_p/\pi-1}\left(u\right)\Lambda_p\left(v\right),\\
\Lambda_p=\phi_{p+1}\left(v\right)~at~\varphi>0,
\Lambda_p=-2p\phi_{p-1}\left(v\right)~at~\varphi<0,\nonumber
\end{eqnarray}
where
\begin{eqnarray}
\gamma_p=\pi|\tan{\psi}|\left(p+\frac{1}{2}\left(1+sign(\varphi)\right)\right).
\end{eqnarray}
The term $I$ corresponds to a wave propagating in the positive
direction along $u$, namely to the incident wave and converted wave.
The term $R$ being proportional to $D_{i\gamma_p/\pi-1}(u)$
corresponds to the reflected wave propagating in the negative
$u$-direction. The coefficients $C_{p}$ in (12) can be chosen in
such a way to fit the incident WKB ordinary wave in the WKB region.
To define $C_p$ we perform the transform (4) of the incident beam
$A\left(x,y\right)\rightarrow
A\left(x,q\right)=\hat{L}\left[A\left(x,y\right)\right]$. Using the
definition of $k$, $u$ and $v$ we express $A (x,q)$ in terms of
these variables $A(x,q)\rightarrow A(u,v)$. We assume $A(u,v)$ at
$u_b\rightarrow -\infty$ in the form
\begin{eqnarray}
A(u,v)=\exp{\left(in_{res}z-i\frac{\cos{\psi}}{2}u_b^2+i\frac{\gamma_p}{\pi}\ln{\left(\sqrt{2\cos{\psi}}u_b\right)}\right)}\tilde{A}(v),\nonumber
\end{eqnarray}
As far as (12) is an expansion in orthogonal functions $\phi_p$, we
can define $C_p$ as follows
\begin{eqnarray}
C_p=\int_{-\infty}^\infty dv \tilde{A}(v)\phi_{p}(v)
\end{eqnarray}
What distinguishes (15) from the coefficients obtained earlier in
~\cite{Irzak a} for the magnetic field with zero shear is the
integrand $\tilde{A}(v)$ which is wider function of $v$ due the
transform (4) than in the case $g=0$. As a consequence of this the
eigen-mode spectrum $C_p$ is changed considerably compare to the
case of the magnetic field with zero shear.
\section{Conversion coefficients}
Since $I$ corresponds to the incident wave at $u<0$ and to the
converted wave at $u>0$ we can derive the conversion coefficient for
the O-X mode conversion using asymptotic expression of
$D_{i\gamma_p/\pi}\left(u\right)$~\cite{Bateman} entering (12). Let
us introduce the conversion coefficient ($T_p$) corresponding to a
separate $p$ term in the sum (12) over the Hermitian polynomials
\begin{eqnarray}
T_p=\exp{\left(-2\pi(p+\mu)\tan{|\psi|}\right)},\mu=1~at~\varphi>0,~\mu=0~at~\varphi<0\\
\tan{|\psi|}=\left(\frac{\sqrt{1+\sin\varphi^2/\left(\cos\varphi-g^2\right)^2}-1}{\sqrt{1+\sin\varphi^2/\left(\cos\varphi-g^2\right)^2}+1}\right)^{1/2}.\nonumber
\end{eqnarray}
As in the case of the magnetic field with zero shear, the conversion
coefficient for the fixed $p$ mode (16) has asymmetry, considered
first time in ~\cite{Gospodchikov2Da}, with respect to the sign of
the angle $\varphi$ between two surfaces $\varepsilon_\parallel=0$
and $\varepsilon_+-n_{res}^2=0$
($\mu=1~at~\varphi>0,~\mu=0~at~\varphi<0$). For the fixed $n_z$ the
conversion coefficient $T$ for the O-X mode conversion may be
expressed as
\begin{eqnarray}
T=\frac{\sum_pT_p\left|C_p\right|^2}{\sum_p\left|C_p\right|^2},
\end{eqnarray}
For the magnetic field with zero shear $g=0$ the transform (4)
yields $A\left(x,y\right)=\hat{L}\left[A\left(x,y\right)\right]$ and
the conversion coefficient of the fixed mode $p$ is performed as
in~\cite{Irzak a}
\begin{eqnarray}
T_p=\exp{\left(-2\pi\tan{\left(|\varphi|/2\right)}\left(p+\mu\right)\right)}.\nonumber
\end{eqnarray}
At $\varphi=0$ with the use of Mehler's formula~\cite{Bateman} we
return to the conversion coefficient
\begin{eqnarray}
T=\frac{\int{dq\left|\tilde{A}(q)\right|^2\exp{\left(-2\pi
q^2/\left(1-g^2\right)^{3/2}\right)}}}{\int{dq\left|\tilde{A}(q)\right|^2}},\\
\tilde{A}\left(q\right)=\int{dy\cdot
\tilde{A}\left(y\right)\exp{\left(-iqy\right)}}\nonumber
\end{eqnarray}
derived firstly by Cairns and Lashmore-Davies
in~\cite{Lashmore-Davies} by the WKB analysis. We remind that $q$
and $g$ are scaled according (2) and $n_z=0$ is assumed. Analyzing
(4), (9), (12) and (14) we can summarize that the magnetic field
shear degrades the efficiency of the O-X mode conversion. An
important feature is increasing of the degradation with increasing
of the mode number. Though the parameter $g$ for the typical
experimental conditions is small $g\sim L_n/L_0\ll 1$ we can expect
remarkable effect of the magnetic field shear on the efficiency of
the O-X mode conversion for the modes with $p\gg 1$.
\section{Conclusion}
In the paper new insight on the 2D O-X mode conversion problem has
been done by considering plasmas confined by the sheared magnetic
field. Ignoring the curvature of the magnetic flux surfaces under an
assumption of high localization of the conversion region and
assuming the straight magnetic field lines we have studied the
problem in the simplest slab geometry accounting for plasma
axisymmetry. Nevertheless, the results being obtained is expected to
hold in more realistic toroidal plasmas. The resulting expressions
for the electric field components (9), (12) and (13) have been
derived that permits finding the conversion coefficient (16)
explicitly. As it has been demonstrated the magnetic field shear
degrades efficiency of the O-X mode conversion. For usual
experimental setup this degradation appears to be remarkable.
Finally, extending the model to the case when the magnetic field
direction do not parallel to the toroidal direction at the O-mode
cut-off surface which is more relevant to tokamak physics needs to
be done. This issue will be studied in future paper.


\section*{References}
\begin{thebibliography}{99} 
\bibitem{Ref1} J~Prienhalter and V~Kopecky  1973 {\it J.
Plasma Physics} {\bf 10} 1
\bibitem{Ref2} H~Weitzner and D~B~Batchelor 1979 {\it Phys. Fluids} {\bf 22} 1355
\bibitem{Ref3} E~Mjolhus 1984 {\it J.Plasma Phys.} {\bf 31} 7
\bibitem{Ref5} M~D~Tokman 1985 {\it Sov. J. Plasma Phys.} {\bf 11} 689
\bibitem{Laqua} H~P~Laqua, V~Erckmann, H~J~Hartfuss et al. 1997 {\it Phys. Rev.
Lett.} {\bf 78} 3467; H~P~Laqua, H~Maasberg, N~B~Marashchenko et al.
2003 {\it ibid.} {\bf 90} 075003
\bibitem{Weitzner} H~Weitzner 2004 {\it Phys.
Plasmas} {\bf 11} 866
\bibitem{Gospodchikov2D} E~D~Gospodchikov, A~G~Shalashov, E~V~Suvorov  2006
{\it Plasma Phys. Control. Fusion} {\bf 48} 869; E~D~Gospodchikov,
A~G~Shalashov 2008 {\it Plasma Phys. Control. Fusion} {\bf 50}
045005; E~D~Gospodchikov, A~G~Shalashov 2008 {\it Phys. Rev. E} {\bf
78} 065602
\bibitem{Irzak}A~Yu~Popov, A~D~Piliya  2007 {\it Plasma Phys.
Reports} {\bf 33} 109-116; A~Yu~Popov 2007 {\it Plasma Phys.
Control. Fusion} {\bf 49} 1599-1610; M~A~Irzak and A~Yu~Popov 2008
{\it Plasma Phys. Control. Fusion} {\bf 50} 025003
\bibitem{Gospodchikov3D} E~D~Gospodchikov, A~G~Shalashov
{\it Proc. Strong Microwaves in Plasmas} July 27 - August 2, 2008,
Nizhny Novgorod.
\bibitem{Lashmore-Davies} R~A~Cairns, C~N.~Lashmore-Davies 2000 {\it
Phys. Plasmas} {\bf 7} 4126
\bibitem{Bateman} Harry Bateman \textit{High transcendental functions}
(MC Graw-Hill Book Company, Inc, 1953)
\end{thebibliography}
\end{document}